\documentstyle[aps,prl,psfig]{revtex}
\begin{document}

\newcommand {\Vy}        {V$_{2-y}$O$_3$}
\newcommand {\V}         {V$_2$O$_3$}
\newcommand {\be}        {\begin{equation}}
\newcommand {\ee}        {\end{equation}}
\newcommand {\ea}        {{\em et al.}}

\twocolumn[\hsize\textwidth\columnwidth\hsize
\csname @twocolumnfalse\endcsname

\title{Combining dynamical mean-field theory and realistic
band structure for \V}
\author{T.~Wolenski, M.~Grodzicki, and J.~Appel\\
Universit\"at Hamburg,
I. Institut f\"ur Theoretische Physik\\
Jungiusstra{\ss}e 9, D-20355 Hamburg.}
\date{\today}
\maketitle  

\begin{abstract}
Recent neutron scattering experiments on \V\ show that the
magnetic fluctuations on the metallic side of the antiferromagnetic 
metal-insulator transition
are not related to the spin structure of the 
insulator, but rather to the bandstructure-driven spin-density
wave phase of the doped system \Vy.
We calculate these magnetic fluctuations starting from a 
Slater-Koster bandstructure and incorporating the correlation
effects through the dynamical mean-field theory (DMFT).
Our results demonstrate that the magnetic properties of the
paramagnetic metallic phase are dominated by the
Fermi surface topology. 
On the other hand, the electron-electron interaction drives 
the paramagnetic metal-insulator transition 
in \V. The transition to the antiferromagnetic insulator by
virtue of orbital ordering is discussed in the framework of the DMFT.
\end{abstract}

\vspace{0.8cm}
]
\narrowtext

\begin{figure}[h]
\hspace{-0.4cm}
\parbox{4cm}{\psfig{file=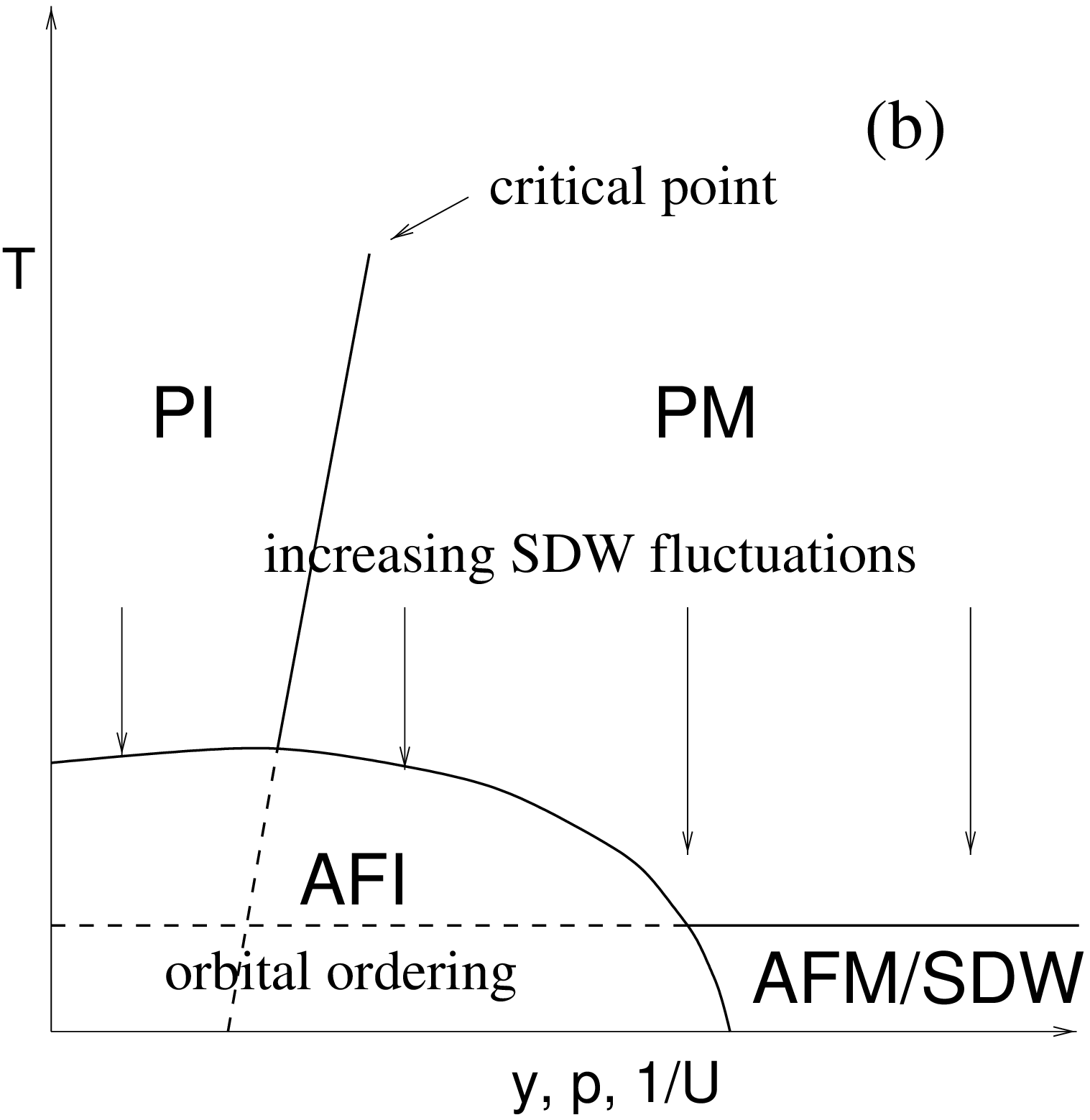,width=4cm}} \hfill
\parbox{4cm}{\psfig{file=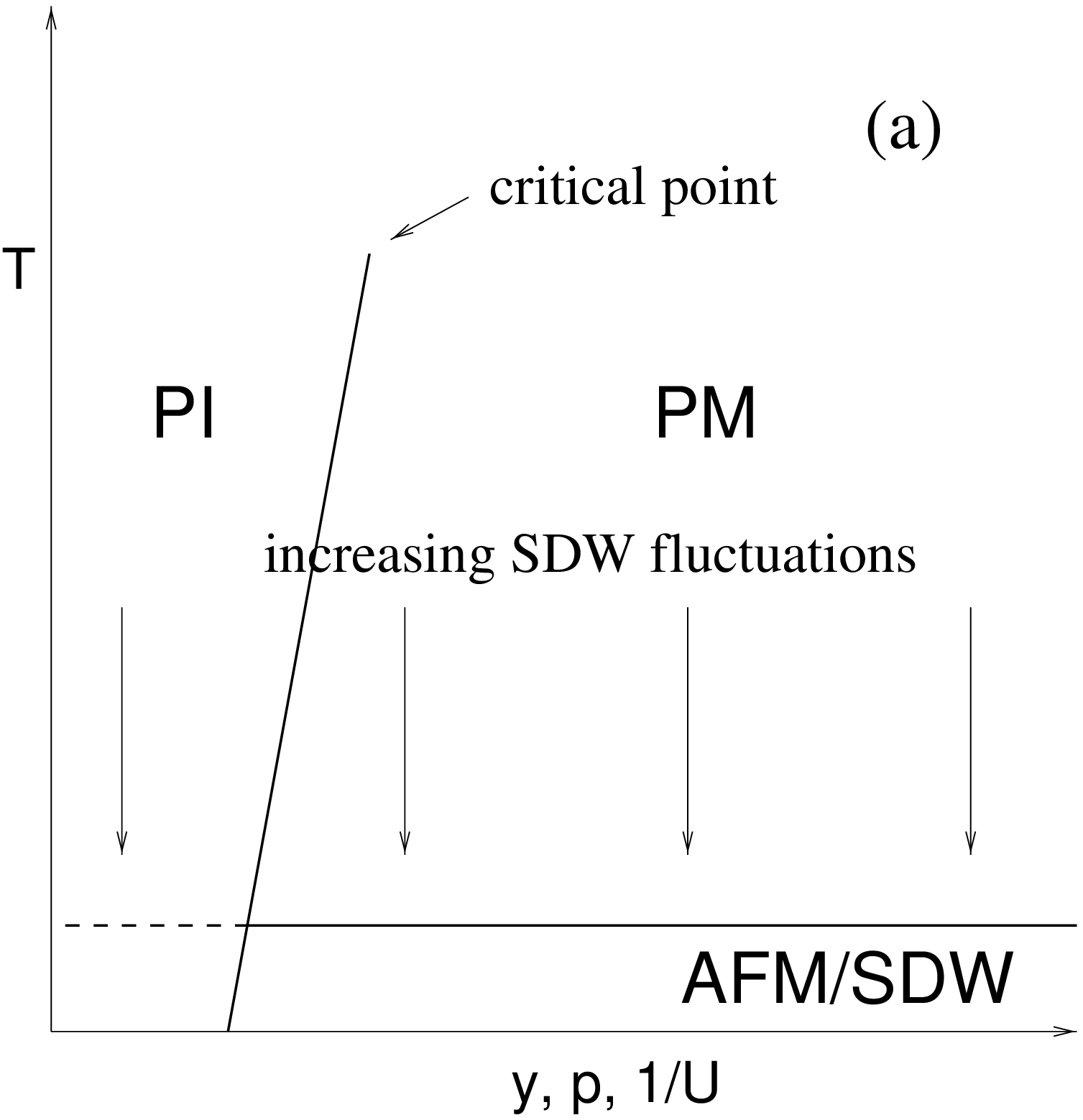,width=4cm}}
\caption{Schematic phase diagram of \V, a) including and b) without the
orbital ordering transition to the AFI. Cf.~text for abbreviations,
Ref.~\protect\cite{WGA1} for numerical values.}
\label{fig:phd}
\end{figure}

Vanadium sesquioxide \V\ is special among the many fascinating 
transition-metal oxides as it served as one of the
first examples of a material exhibiting an
interaction-driven Mott-Hubbard metal-insulator transition (MIT). 
The phase diagram of \V\ (cf.~Fig.~\ref{fig:phd}~a) shows four different
phases. The controlling variable, besides the temperature, is the
ratio of local Coulomb interaction $U$ and bandwidth $D$ as evidenced by the
similar response to substitutional doping with Cr and Ti or 
vanadium deficiency on the one hand and the application 
of hydrostatic pressure on the other hand.
At low temperatures, an antiferromagnetic insulator (AFI)
and an antiferromagnetic metal (AFM) occur. The spin structures
of these magnetically ordered phases differ dramatically \cite{Bao_SDW}.
At higher temperatures, a first-order paramagnetic 
metal-insulator transition occurs, ending in a second-order 
critical point. The magnetic fluctuations in the paramagnetic phases
have recently been shown to correspond to the long-range order
observed in the AFM \cite{Bao_fluct}.

Experimentally, there are indications that the transition to the 
AFM phase is a spin-density wave transition, i.e.\
an electronic instability due to the formation of electron-hole pairs
because the Fermi surface has nesting sheets: Parts
of the Fermi-surface are almost parallel, providing large phase space of 
low-energy electron and hole excitations at positions ${\bf k}$ and
${\bf k+Q}$ in momentum space that are connected by the same 
nesting vector ${\bf Q}$. We recently demonstrated that this
mechanism indeed provides an explanation for the properties of
the AFM phase \cite{WGA1}.


The point of departure is the electronic structure of \V.
Early attempts to calculate the bandstructure of \V\ \cite{band_early}
were based on tight-binding calculations in a strongly reduced
basis set, and are thus at best qualitatively correct. Recently,
state-of-the-art density functional theory (DFT)
results were made available \cite{Mattheiss}.
To incorporate these results we
employ the Slater-Koster method, which uses the 
DFT energy eigenvalues at points of high symmetry to parameterize
tight-binding matrix 
\noindent
elements, which can then be evaluated
at any point in the Brillouin zone. This reduces the numerical effort
to obtain reliable bandstructure information. Details of this approach
are provided in a recent article \cite{WGA1}.

Using this bandstructure we are able to calculate the properties
of the SDW transition \cite{WGA1} within a weak-coupling Hartree-Fock
mean-field theory, and find excellent agreement
with the experimental situation \cite{Bao_SDW,Pieper}. The value
of the Fermi surface nesting vector responsible
for the spin-density wave instability is found to be
${\bf Q}=1.69 \, c^*$, the transition temperature to be 10~K
at an effective Hubbard interaction $U_c \simeq 0.5$~eV,
in good agreement with 
optical experiments \cite{Thomas}.
The success of the weak-coupling theory shows that the
spin-density wave is Fermi-surface-driven, and is not a frustration
phenomenon of the antiferromagnetic insulator as suggested
previously \cite{frust_SDW,Rozenberg}. The occurrence of the paramagnetic MIT
on the other hand indicates that correlations are important
in \V. 


The dynamical mean-field theory (DMFT) employed here is based on
the insight that in the limit of high spatial dimensions or,
better, for high lattice co-ordination spatial fluctuations
become irrelevant, while dynamical quantum fluctuations remain 
important as shown by Vollhardt and Metzner \cite{Metzner}. 
This procedure allows to chose one lattice site $i=0$ and 
integrate out all other
degrees of freedom, leading to a local effective action
$S_{\text{eff}}[{\cal G}_0]$, which is a functional of an effective field
${\cal G}_0(\tau-\tau')$. The effective field describes the interaction
of the local problem at $i=0$ with a bath made up of the rest
of the lattice. An electron at $i=0$ can go into the bath at
(imaginary) time $\tau$ and return to the local site at $\tau'$.
Excellent reviews of the DMFT include
Ref.~\cite{DMFT_reviews}.

The solution of the local problem remains non-trivial. Here, we use
the so-called modified iterated perturbation theory (MIPT), 
which becomes exact in both the weak coupling and atomic limit
and at low and high frequencies \cite{MIPT}.
The self-energy $\Sigma(i\omega_n)$ in MIPT is obtained through
the Ansatz
\be
\Sigma^{\text{MIPT}}(i \omega_n)=U\left<n\right> + 
\frac{a \Sigma^{(2)}(i \omega_n)}{1-b \Sigma^{(2)}(i \omega_n)},
\ee
where $\Sigma^{(2)}(i \omega_n)$ is the second-order perturbation
theory result,
\be
\Sigma^{(2)}(i \omega_n) =  
U^2 \!\! \int_0^\beta \! d\tau \, e^{i \omega_n \tau} {\cal G}_0 (\tau)^3.
\ee
Details on obtaining $a$ and $b$ to fit
various limiting case are given in Ref.~\cite{MIPT}.
The lattice enters through a self-consistency condition: The effective
field of the local problem is identified with the local 
Green's function of the lattice problem, and the full lattice Green's
function $G({\bf k},i \omega_n)$ is calculated from the local
self-energy. The self-consistency equation is 
\be
\Sigma(i\omega_n)={\cal G}_0^{-1}(i\omega_n)-G^{-1}(i\omega_n)
\ee
and
\be
G(i\omega_n)=\sum_{\bf k} \! G_{\bf k} (i\omega_n)
=\sum_{\bf k} \frac{1}{i\omega_n-\epsilon_{\bf k}+\mu-\Sigma(i\omega_n)}.
\label{eq:G}
\ee
The momentum summation brings in the density of states of the
actual lattice.
Before employing the MIPT, we 
point out that
Rozenberg \ea\ \cite{Rozenberg} applied the
DMFT on a Bethe
lattice to get a fairly complete picture
of the interaction-driven metal-insulator transition, where they
introduce phenomenological frustration to suppress a Slater-type
transition.
A first central point here is that we show that their results 
regarding the paramagnetic MIT
essentially carry over to the case of \V\ including the realistic
bandstructure as presented in Ref.~\cite{WGA1}.

\begin{figure}[h]
\hspace{0.5cm} \psfig{file=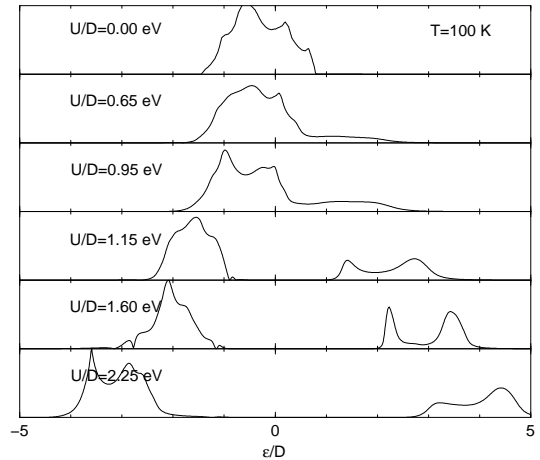,width=7cm}
\caption{Development of the spectral function as the interaction $U$
is tuned through the MIT.}
\label{fig:spectral}
\end{figure}

In Fig.~\ref{fig:spectral} we show the development of the spectral
function $A(\omega)=-1/\pi \, G(\omega+i0^+)$ as the 
interaction $U$ is tuned through the MIT. 
It is obtained after solving the DMFT in the MIPT approximation
and Pad\'e-transforming the results to the real frequency axis.
The transfer of spectral weight to higher frequencies and the
eventual formation of upper and lower Hubbard bands is clearly seen.
As long as the system stays metallic $A(0)$ does not change, as
required for a local self-energy. The narrowing of the 
quasi-particle peak at $\omega=0$ can be 
\begin{figure}[h]
\center{\psfig{file=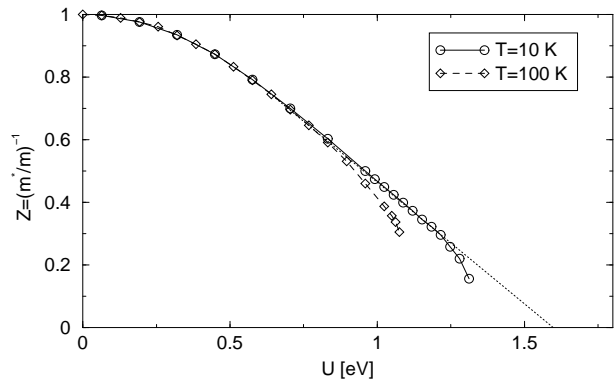,width=8cm}}
\caption{Divergence of the effective mass $m^*$ at the MIT. Dotted line
is an extrapolation to $T=0$.}
\label{fig:Z}
\end{figure}
\noindent
described as a divergence of the effective mass $m^*$ or reduction of
the quasiparticle weight $Z$, given by
\be
\frac{m^*}{m}=\frac{1}{Z}=1-\frac{\partial}{\partial \omega} \, 
\mbox{Re} \, \left. \Sigma(\omega+i 0^+) \right|_{\omega=0}.
\ee
The result is shown in Fig.~\ref{fig:Z}. The finite temperature
curves extrapolate to a linear behavior at $T=0$.

\begin{figure}[h]
\center{\psfig{file=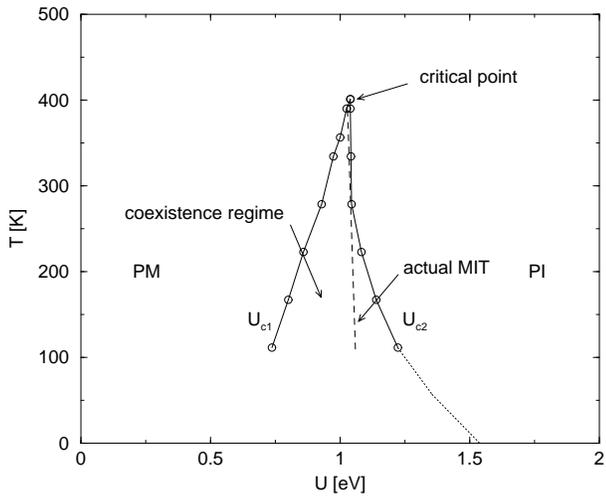,width=8cm}}
\caption{Stability regime of metallic and insulating solutions to the
DMFT equations. Dashed line indicates physical phase transition.}
\label{fig:metins}
\end{figure}

Rozenberg \ea\ found a coexistence regime for metallic and insulating
solutions of the DMFT equations. We confirmed this finding for
the realistic bandstructure of \V. Fig.~\ref{fig:metins} shows the
stability regimes, the metallic solution is stable for $U< U_{c2}$,
the insulating solution for $U>U_{c1}$. The physical phase transition
occurs at the crossing of the free energy surfaces of these
curves, indicated by the dashed line in Fig.~\ref{fig:metins}.
The dotted line is an extrapolation of $U_{c2}$ to $T=0$
based on Fig.~\ref{fig:Z}. The values of the transition 
temperatures and the effective interaction agree well with 
experimental values \cite{Thomas}.

While the consideration of the realistic bandstructure did only 
provide quantitative corrections of the paramagnetic MIT, 
it is essential for obtaining even a qualitative picture of the 
magnetic properties. Rozenberg \ea\ extended their study to include
frustrated magnetism in 
\noindent
the Bethe lattice, but this approach is
not able to explain the different spin structures of the AFI and
AFM and the absence of any AFI-type fluctuation in the PM in 
close proximity to the AFI \cite{Bao_fluct}.

Since the magnetic structure of the AFM phase
and the corresponding fluctuations in the paramagnetic phases are 
experimentally known to be incommensurate, the only way to study
them is through the dynamic spin susceptibility, 
as no precursors of a symmetry breaking phase can be seen on 
the one-particle level as $d \rightarrow \infty$ \cite{MuellerHartmann}. 
The DMFT is used here as an approximation to calculate the momentum
dependent susceptibility; it is determined by the
lattice Green's functions, Eq.~(\ref{eq:G}), as
\be
\chi^0({\bf q},\omega)=
-\sum_{{\bf k},\nu} G({\bf k},\nu) G({\bf k}+{\bf q},\omega+\nu).
\label{eq:chi_0}
\ee
The full susceptibility $\chi$ is obtained from $\chi^0$ after
the vertex functions are taken into account \cite{DMFT_reviews}, 
which become momentum independent as $d \rightarrow \infty$.

\begin{figure}[h]
\center{\psfig{file=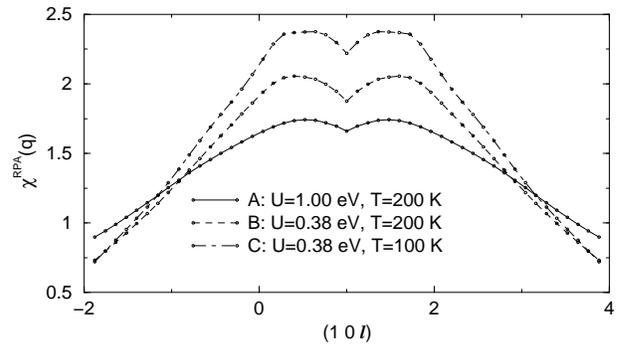,width=8cm}}
\caption{Dynamic susceptibility at different points in the phase diagram
(cf.~text). The wave vector at the maximum of $\chi({\bf q},\omega)$ 
corresponds to the long-range order observed in the AFM.}
\label{fig:chiq}
\end{figure}

The dynamic susceptibility is connected with the neutron scattering
intensity through the fluctuation-dissipation theorem, which
allows the comparison with the measurements of 
Bao \ea\ \cite{Bao_fluct,Bao_review}. We present 
three characteristic points in the PM phase:
A is chosen close to the paramagnetic MIT and close to the AFI,
B and C are at different temperatures above the AFM low-temperature
phase. In Fig.~\ref{fig:chiq} we show the momentum dependence of 
the susceptibility for a given small frequency $\omega$.
At all temperatures the maximum in the susceptibility is at the 
wave vector ${\bf Q}$ corresponding to the AFM spin-density wave structure.
Remarkably, at A no evidence of the close AFI phase is seen, in agreement
with neutron scattering results. The comparison of B and C shows
the increase of the intensity at ${\bf Q}$ as the spin-density wave
transition is approached.

\begin{figure}[h]
\center{\psfig{file=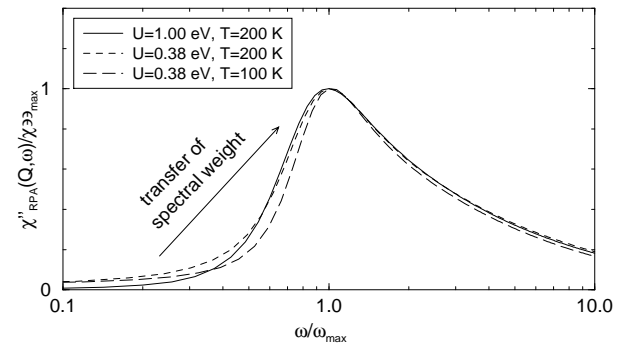,width=8cm}}
\caption{Frequency dependence of the dynamic susceptibility at the
same parameter values as in Fig.~\protect{\ref{fig:chiq}}.}
\label{fig:chiw}
\end{figure}

In Fig.~\ref{fig:chiw} the
frequency dependence of $\chi({\bf q}={\bf Q},\omega)$ is 
\noindent
shown, scaled 
approximately on a single curve. This is to be compared with the 
results of Bao \ea\ \cite{Bao_review}. Bao \ea\  
fit their results with the SCR theory for itinerant antiferromagnetism
\cite{SCR}. Our curve has spectral weight transferred to higher
frequencies compared to the SCR due to the strong correlations,
and does improve on the fit to the experimental result.

We demonstrate that the combination of the DMFT
with the realistic bandstructure provides a natural
description of the paramagnetic metal-insulator transition,
the spin-density wave transition, and the magnetic fluctuations of the 
paramagnetic phases. This is summarized in Fig.~\ref{fig:phd}~b).
In particular, this parameter-free calculation gives good
quantitative agreement with experimental results, both for the
spin-density wave \cite{WGA1} and the MIT.
The absence of any AFI-type fluctuations outside the AFI as 
observed in neutron scattering experiments is due to the strongly
first order transition to the AFI.


Let us now turn to the AFI. Castellani \ea\
\cite{Castellani} suggested several years ago, that the essential
mechanism of the AFI is orbital ordering \cite{Kugel}. In \V,
the vanadium $e_g$ orbitals form a doubly degenerate, 
quarter-filled band. The degeneracy is lifted in second-order
perturbation theory, leading to a favoring
of ferromagnetic spin and ``antiferromagnetic orbital'' coupling,
i.e.\ a staggered occupation of the two degenerate orbitals.
\begin{figure}[h]
\hspace{2cm}\psfig{file=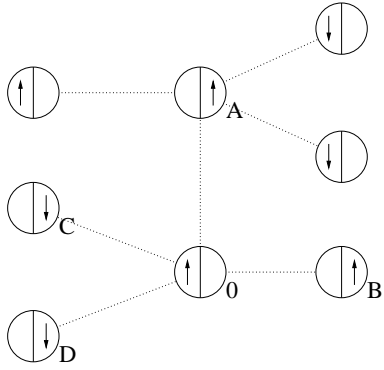,width=5cm}
\vspace{0.2cm}
\caption{Proposed cluster of vanadium sites to study orbital ordering
in the AFI of \V. The split circles denote the occupation of the
two degenerate orbitals at each site \protect\cite{Castellani,Kugel}.}
\label{fig:cluster}
\end{figure}
\noindent
Castellani \ea\ were able to apply this idea to \V\ and demonstrate
that the observed spin structure of the AFI can be explained
in this approach.
It is thus necessary to go beyond the effective
one-band model studied so far \cite{WGA1}.

While this complements our picture of \V, the approach 
of Castellani has two shortcomings from todays point
of view: The authors had to rely on the bandstructure calculations
available at the time \cite{band_early} and strong correlations
have not been taken into account, even in a regime where the
paramagnetic MIT is known to occur. These corrections 
are not expected to be essential, yet for a complete picture of \V\
it is desirable to include them in the calculation.
To that end, the DMFT equations have to be generalized to a
cluster of vanadium sites as shown in Fig.~\ref{fig:cluster}.
This approach is similar to introducing two sublattices to
study simple AB-type antiferromagnetism. Here, four neighbor 
sites of $i=0$ have to be considered. The Green's
functions for these sites are related by
\be
G^0_{\nu \sigma}
=G_{\overline{\nu}\sigma}^A
=G_{\overline{\nu}\sigma}^B
=G_{\overline{\nu}\overline{\sigma}}^C
=G_{\overline{\nu}\overline{\sigma}}^D
\ee
where $\nu$ is an orbital index and we suppressed the argument $i\omega_n$. 
The orbitally dependent DMFT
equations have been derived by Momoi and Kubo and by
Held and Vollhardt \cite{MomoiHeld} for the doubly-degenerate 
Hubbard model. The hopping matrix elements $t_{ij}^{\nu\nu'}$
and the eigenvectors $A_{\nu \alpha}({\bf k})$, 
where $\alpha$ is a band index,
can be obtained from the bandstructure calculation.
One has
\be
t_{ij}^{\nu\nu'}=\sum_{\bf k} e^{i{\bf k} \cdot ({\bf R}_i-{\bf R}_j)}
\sum_\alpha A_{\alpha\nu}({\bf k}) \epsilon_{\bf k}^\alpha 
A_{\alpha\nu'}^*({\bf k}).
\ee
To solve the coupled problem is straightforward in principle
but has not yet been performed for \V.

In conclusion, the various phase transitions of \V\ are consistently
explained once the physics of strong correlation and the
realistic bandstructure are combined. Apart from the
calculation of orbital ordering within DMFT, some other open
questions remain. A systematic $1/d$-expansion would complement the
assumption of high spatial dimensions in the treatment of the
strong-correlation problem. The importance of disorder, 
which is certainly present at least in the doped \V\ compounds,
has not been studied. Dobrosavlevi\'c and Kotliar \cite{Dobrosavlevic}
provided a promising extension of the DMFT to the dirty limit. 
And, finally, the
absence of superconductivity in \V\ is still a challenging problem. 
The solution to this last problem promises new insights into
the basic mechanism of high-temperature superconductivity in the
cuprates. One possible scenario is that the large local moments 
known to occur in metallic \Vy\ \cite{Pieper} destroy the superconductivity,
so that it would be interesting to look for superconductivity
in completely stoichiometric \V, where the metallic phase has 
been stabilized through application of hydrostatic pressure.

T. W. acknowledges financial support from the Universit\"at Hamburg;
this work is in partial fulfillment of his doctoral degree at
this university.
%


\vspace{-0.5cm}
\begin{thebibliography}{99}
\vspace{-1.2cm}
\bibitem{Bao_SDW}
W. Bao {\em et al.}, {\em Phys.\ Rev.\ Lett.} {\bf 71}, 766 (1993).

\bibitem{Bao_fluct}
W. Bao \ea,
{\em Phys. Rev. B} {\bf 54}, R3726 (1996); W. Bao {\em et al.},
{\em Phys. Rev. Lett.} {\bf 78}, 507 (1997).

\bibitem{WGA1}
T.~Wolenski, M.~Grodzicki, and J.~Appel, 
{\em Phys.\ Rev.\ B} {\bf 58}, 303 (1998).

\bibitem{band_early}
I. Nebenzahl and M. Weger, {\em Phil.\ Mag.} {\bf 24}, 1119 (1971);
J.~Ashkenazi and T.~Chuchem, {\em Phil.\ Mag.} {\bf 32}, 763 (1975).

\bibitem{Mattheiss}
L.\ F.\ Mattheiss, {\em J. Phys.: Condens.\ Matter} {\bf 6}, 6477 (1994);
M.~Grodzicki and O.~Jepsen, unpublished.

\bibitem{Pieper}
S. Langenbuch, M. W. Pieper, P.~Metcalf, and J.~M.~Honig,
{\em Phys.\ Rev.\ B} {\bf 53}, R472 (1996).

\bibitem{Thomas}
G. A. Thomas \ea,
{\em Phys. Rev. Lett.} {\bf 73}, 1529 (1994).

\bibitem{frust_SDW}
S.~Sachdev and A.~Georges, {\em Phys.\ Rev.\ B} {\bf 52}, 9520 (1995).

\bibitem{Rozenberg}
M.~J.~Rozenberg, X.~Y.~Zhang, and G.~Kotliar,
{\em Phys.\ Rev.\ Lett.} {\bf 69}, 1236 (1992);
M.~J.~Rozenberg, G.~Kotliar, and X.~Y.~Zhang,
{\em Phys.\ Rev.\ B} {\bf 49}, 10181 (1994).

\bibitem{Metzner}
W.~Metzner and D.~Vollhardt, {\em Phys.\ Rev.\ Lett} {\bf 62}, 324 (1989);
W.~Metzner, {\em Phys.\ Rev.\ B} {\bf 43}, 8549 (1991).

\bibitem{DMFT_reviews}
A. Georges, G. Kotliar, W. Krauth, and M.~J.~Rozenberg,
{\em Rev. Mod. Phys.} {\bf 68}, 13 (1996); 
D.~Vollhardt, in {\em ``Correlated Electron Systems,''} Ed.~V.~J.~Emery,
World Scientific, Singapore, 1993.

\bibitem{MIPT}
H.~Kajueter and G.~Kotliar, {\em Phys.\ Rev.\ Lett.} {\bf 77}, 131 (1996);
M.~Potthoff, T.~Wegner, and W.~Nolting, 
{\em Phys.\ Rev.\ B} {\bf 55}, 16132 (1997). 

\bibitem{MuellerHartmann}
E.~M\"uller-Hartmann, {\em Z.\ Phys.\ B} {\bf 76}, 211 (1989).

\bibitem{Bao_review}
W.~Bao \ea,
{\em Phys. Rev. B} {\bf 58} (19) (1998), cond-mat/9804320.

\bibitem{SCR}
K.~Nakayama and T.~Moriya, {\em J.\ Phys.\ Soc.\ Jap.} {\bf 56}, 2918 (1987).

\bibitem{Castellani}
C.~Castellani, C.R.~Natoli, and J.~Ranninger, {\em Phys.\ Rev.\ B}
{\bf 18}, 4945 (1978); {\em ibid.}, {\em Phys.\ Rev.\ B}
{\bf 18}, 4967 (1978).

\bibitem{Kugel}
K.I.~Kugel and D.I.~Khomskii, {\em Sov.-Phys.\ JETP} {\bf 37}, 725 (1973);
{\em Sov.-Phys.\ Usp.} {\bf 25}, 231 (1982);
M.~Cyrot and C. Lyon-Caen, {\em J.\ Physique} {\bf 36}, 253 (1975).

\bibitem{MomoiHeld}
T.~Momoi and K.~Kubo, {\em Phys.\ Rev.\ B} {\bf 58}, R567 (1998);
K.~Held and D.~Vollhardt, cond-mat/9803182.

\bibitem{Dobrosavlevic}
V.~Dobrosavlevi\'c and G.~Kotliar, {\em Phys.\ Rev.\ Lett.}
{\bf 78}, 3943 (1997); cond-mat/9706231.

\end{thebibliography}
\end{document}